\documentclass[prd,nofootinbib,preprint,superscriptaddress,twocolumn,10pt]{revtex4}
\pdfoutput=1
\usepackage[T1]{fontenc}
\usepackage{amsmath,amssymb}
\usepackage{physics}
\usepackage{braket}
\usepackage{epsfig}
\usepackage{graphicx}
\usepackage[usenames,dvipsnames]{color}
\usepackage{subfigure}
\usepackage{slashed}
\usepackage[normalem]{ulem}
\usepackage[colorlinks,citecolor=blue]{hyperref}

\definecolor{nicegreen}{rgb}{0.08,0.7,0.1}

\usepackage{color}
\usepackage{comment}
\usepackage{tikz-feynman}
\usepackage{amsfonts}
\DeclareMathSymbol{\shortminus}{\mathbin}{AMSa}{"39}
\begin{document}


\title{Reviving WIMP dark matter with temperature-dependent couplings}

\author{Debasish Borah}
\email{dborah@iitg.ac.in}
\affiliation{Department of Physics, Indian Institute of Technology Guwahati, Assam 781039, India}

\author{Arnab Dasgupta}
\email{arnabdasgupta@pitt.edu}
\affiliation{Pittsburgh Particle Physics, Astrophysics, and Cosmology Center, Department of Physics and Astronomy, University of Pittsburgh, Pittsburgh, PA 15260, USA}

\author{Tong Arthur Wu}
\email{tow39@pitt.edu}
\affiliation{Pittsburgh Particle Physics, Astrophysics, and Cosmology Center, Department of Physics and Astronomy, University of Pittsburgh, Pittsburgh, PA 15260, USA}

\preprint{PITT-PACC-2508}
\preprint{CETUP2025-010}

\begin{abstract}
The persistent null results at dark matter (DM) direct-detection experiments have pushed the popular weakly interacting massive particle (WIMP) DM to tight corners. Generic WIMP models with direct-detection rate below the current upper limits often lead to a thermally overproduced relic abundance after freeze-out. To resolve this conundrum, we propose a novel scenario where DM has temperature-dependent couplings with the standard model (SM) bath. A scalar field having a large vacuum expectation value (VEV) at high temperatures generates sizeable DM-SM interactions leading to efficient DM annihilations responsible for generating the desired thermal relic. At lower temperatures, the scalar field VEV settles down to a small value as a result of a phase transition which can generically be of first order, effectively leading to suppressed DM-SM interaction rate at low temperature, consistent with null results at direct-detection experiments. Upper bound on thermal DM mass forces the first-order phase transition (FOPT) to occur at scales such that the corresponding gravitational wave signal remains within reach of future experiments like LISA.
\end{abstract}
\maketitle
{\it Introduction}: As suggested by observations from astrophysics and cosmology related experiments, the matter content of our present Universe is dominated by dark matter (DM) contributing approximately five times of ordinary baryonic matter \cite{Planck:2018vyg, ParticleDataGroup:2024cfk, Cirelli:2024ssz}. While the standard model (SM) of particle physics does not have any particle DM candidate, several beyond standard model (BSM) proposals have been put forward to explain the origin of DM. Among them, the weakly interacting massive particle (WIMP) \cite{Kolb:1990vq, Jungman:1995df, Bertone:2004pz} has been the most widely studied one. A typical WIMP, by virtue of its sizeable non-gravitational interactions with the SM bath, can be thermally produced in the early Universe with its relic set by thermal freeze-out. The same DM-SM interactions can also lead to observable DM-nucleon scattering at terrestrial detectors. However, null results at direct-detection experiments like \texttt{LUX-ZEPLIN (LZ)} ~\cite{LZ:2024zvo}, \texttt{XENONnT} \cite{XENON:2025vwd}, \texttt{PandaX-4T} \cite{PandaX:2024qfu} have already ruled out a large part of the parameter space of the simplest WIMP models.

While it is possible to keep DM direct-detection rate suppressed while being consistent with the thermal relic criteria, this requires specific couplings and masses to keep the DM-nucleon scattering either momentum suppressed or accidentally small \cite{Hill:2011be, Hill:2013hoa, Hisano:2011cs, Chen:2019gtm, Arcadi:2025ifl}. However, in the absence of these, generic WIMP models with couplings required to keep direct-detection rate below the current upper bound lead to thermal overproduction. In this letter, we propose a novel framework to revive WIMP parameter space without relying on a tuned parameter space or specific Lorentz structure of DM-SM couplings. We consider a temperature (T)-dependent DM-SM coupling where the vacuum expectation value (VEV) of a scalar field $\eta$ dictates the strength of the coupling. Although the VEV of $\eta$ remains large at the epoch of DM freeze-out, keeping the DM annihilation rate in the typical WIMP ballpark, it gets restored to a small value at lower temperatures leading to suppressed DM-SM coupling consistent with the small direct-detection rate. While DM detection prospects in such a setup remain subdued, consistent with null results, the VEV restoration of the scalar field $\eta$ can be a strong first-order phase transition (FOPT) with observable signatures at gravitational waves (GW) experiments. Several earlier works \cite{Cohen:2008nb, Cui:2011qe, Baker:2016xzo, Baker:2017zwx, Bian:2018bxr, Baker:2018vos, Bian:2018mkl, Heurtier:2019beu, Croon:2020ntf, Elor:2021swj, Hashino:2021dvx, Bian:2021dmp, Adhikary:2024btd, Allahverdi:2024ofe} also studied the role of FOPT on DM relic, but they were limited to studying the impact of FOPT on masses of DM or mediators of DM-SM interactions only. In another work \cite{Hektor:2018esx}, a two-step electroweak phase transition was used to decouple the processes responsible for generating DM relic from the ones producing indirect-detection signals. Instead of considering FOPT effects on masses or on annihilation channels of a specific model, we adopt a model-independent approach where a phase transition is responsible for generating temperature-dependent couplings of DM with the SM bath\footnote{Changes in DM-SM couplings at different scales have also been studied within specific models incorporating running of couplings \cite{Sannino:2014lxa} and varying kinetic mixing \cite{Gan:2023wnp}.}. In our setup, we consider, for the first time, the impact of T-dependent DM-SM couplings in keeping DM detection rates suppressed at terrestrial experiments while still allowing a large annihilation rate in the early Universe.

We consider an effective field theory (EFT) setup to parametrize DM-SM interactions and constrain the cut-off scale $\Lambda$, DM mass, as well as the VEV of $\eta$ consistent with the desired WIMP phenomenology. While VEV restoring transition can occur at any scale above the big bang nucleosynthesis (BBN) epoch \cite{Bai:2021ibt}, the unitarity upper limit on DM mass $m_{\rm DM} \lesssim \mathcal{O}(100\, \rm TeV)$ \cite{Griest:1989wd} forces the nucleation to occur at a scale $T < T_f \lesssim \mathcal{O}(10\, \rm TeV)$ below DM freeze-out $(T_f)$ such that DM freeze-out relic remains unaffected. This keeps the GW spectrum from first-order VEV restoring phase transition within near-future experiments like LISA. Such a low-scale transition also keeps the new scalar degrees of freedom associated with $\eta$ within reach of terrestrial detectors like the large hadron collider (LHC). \\

{\it The framework}: Without any loss of generality, we adopt an EFT approach\footnote{See earlier works \cite{Beltran:2008xg, Cao:2009uw, Goodman:2010ku, Beltran:2010ww, Fitzpatrick:2012ix} on DM EFT, also summarised in a recent review \cite{Bhattacharya:2021edh}.} to describe DM-SM interactions. We provide two working benchmark examples of DM EFT: (i) Dirac fermion singlet DM $\chi$ with Higgs portal interactions (BM1) \cite{Fedderke:2014wda}, (ii) scalar singlet DM $\phi$ with effective couplings to SM fermions $f$ (BM2) \cite{Cao:2009uw}. While these operators can arise at lower dimensions, we consider dimension-six and dimension-seven operators which also include an additional scalar $\eta$ whose VEV is responsible for T-dependent DM interactions. The corresponding interaction Lagrangian can be written as
\begin{equation}
  -\mathcal{L}_{\rm DM-SM}=\begin{cases}
        \frac{1}{\Lambda^2} \eta \bar{\chi}\chi H^\dagger H ,& \hspace{0.5em}{\rm fermion \, DM} \\[1mm]
        \frac{1}{\Lambda^3} \eta \phi \phi H \bar{f} f ,& \hspace{0.5em}{\rm scalar \, DM} 
    \end{cases} \label{eq:L}
\end{equation}
where $H$ denotes the SM Higgs doublet. In addition to SM gauge invariance, the effective operators are also invariant under a $Z_2$ symmetry, responsible for keeping DM stable. Additional symmetries like a softly broken $Z_3$ symmetry acting non-trivially on DM and $\eta$, may be imposed to ensure that the above operators are responsible for leading DM-SM interactions\footnote{Irrespective of additional symmetries, scalar DM can always have renormalisable Higgs portal interactions, which we assume to be sub-dominant.}. It is also possible to have softly broken global $U(1)$ symmetry responsible for ensuring stability of DM by virtue of a remnant $Z_2$ symmetry while also ensuring DM-SM operators given in Eq. \eqref{eq:L} as the leading ones. In addition to such symmetry realizations, one can also find different possible UV completions of these operators. For example, a heavy scalar $\xi$ coupling to fermion DM as $\xi \overline{\chi_L} \chi_R$ and to scalars as $\xi^\dagger \eta H^\dagger H$ can result in dimension-six fermion DM operator in Eq. \eqref{eq:L} at low energy after integrating out heavy scalar degrees of freedom associated with $\xi$. A more explicit UV completion is given in Appendix \ref{appen3}. We remain agnostic about such symmetry realizations as well as UV completion of these operators in this work. As pointed out earlier, a non-zero VEV of $\eta$ at high temperature leads to sizeable DM-SM coupling while its VEV restoration at low temperature is responsible for feeble DM-SM interaction at terrestrial experiments.

In order to realize the desired VEV profile of $\eta$ independent of electroweak symmetry breaking, we consider another singlet scalar $\sigma$ having sizeable coupling to $\eta$ with the relevant scalar potential given by
\begin{align}
    V(\sigma, \eta) & = \frac{\mu^2_\sigma}{2} \sigma^2 + \frac{\lambda_\sigma}{4} \sigma^4 -\frac{\mu^2_\eta}{2} \eta^2 +\frac{\lambda_\eta}{4} \eta^4 \nonumber \\ 
    & + \frac{\lambda_{\sigma\eta}}{4} \sigma^2 \eta^2 - \tilde{\mu} \sigma^2 \eta.
    \label{eq:scalpot}
\end{align}
In order to study the finite-temperature behavior of the potential, one also has to include the Coleman-Weinberg correction $V_{\rm CW}$\cite{Coleman:1973jx} together with the thermal correction $V_{\rm th}$ \cite{Dolan:1973qd,Quiros:1999jp}. For better accuracy, we use the dimensional reduction method \cite{Ginsparg:1980ef,Kajantie:1995dw,Cline:1997bm,Ekstedt:2022bff} to compute the finite-temperature effective potential of $\eta, \sigma$ at two-loop level, the details of which are given in Appendix \ref{appen1}. 

\begin{figure}[h]
    \centering
    \includegraphics[width=0.5\linewidth]{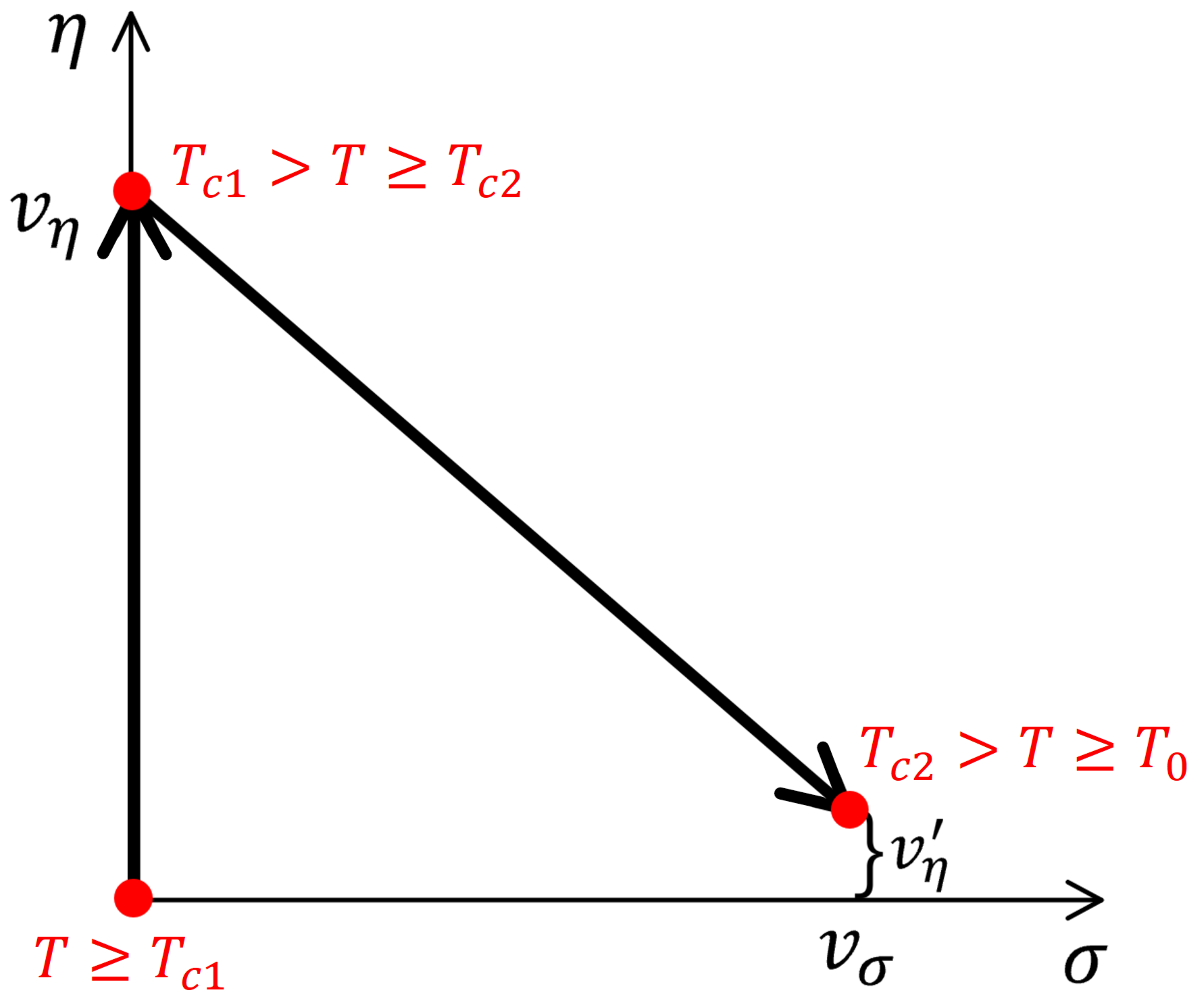} 
    \caption{The evolution of the scalar VEVs in the two-step phase transition.}
    \label{fig:vev2D}
\end{figure}

We consider a two-step phase transition such that the VEVs of $(\eta, \sigma)$ follow $(0,0) \rightarrow (v_\eta, 0) \rightarrow (v'_\eta, v_\sigma)$ with $v'_\eta \ll v_\eta,v_\sigma$. The small VEV of $\eta$ at present temperature $T =T_0$ is realized by a small trilinear interaction term in the scalar potential given by Eq. \eqref{eq:scalpot}. The desired trajectory of the VEVs is shown in Fig. \ref{fig:vev2D}. Initially, at $T > T_{c1}$, the VEVs of $\sigma$ and $\eta$ lie at $(0,0)$. At $T_{c1}> T > T_{c2}$, $\eta$ first acquires a large VEV while $\sigma$ remains at zero. Subsequently, at $T_{c2}>T$, $\sigma$ gets a large VEV while the VEV of $\eta$ almost vanishes up to $v_{\eta}'\propto \tilde{\mu}$ from the small trilinear term. Such a configuration remains till $T=T_0$, the present temperature. We also ensure that DM freezes out after the first transition whereas the second transition occurs after DM freeze-out, resulting in vanishing or small effective DM-SM coupling. If we consider a more minimal setup with SM Higgs playing the role of $\sigma$, it requires a higher $T_{c1}$ and therefore a larger $m_{\rm{DM}}$ above the electroweak scale \cite{Hektor:2018esx}. In order to cover the entire allowed mass range of thermal DM, we therefore consider $\sigma$ to be an additional singlet scalar.  \\

{\it Dark Matter Phenomenology}: The relic abundance of DM can be estimated by solving the appropriate Boltzmann equation. For a typical WIMP DM, the thermal-averaged annihilation cross-section of DM into SM particles govern its freeze-out relic.

For fermion DM, the $v_\eta$-dependent $\chi \bar{\chi} \rightarrow f\bar{f}$ annihilation is enhanced near the Higgs resonance, while the $v_\eta$-independent $\chi \bar{\chi} \rightarrow \eta h$ remains subleading. Here $h$ denotes the physical SM Higgs boson. For scalar DM, the dominant annihilation channels are $\phi \phi \rightarrow f \bar{f}$ which are controlled by $v_\eta$, while $2 \rightarrow 3$ processes like $\phi \phi \rightarrow \eta f \bar{f}$ remain phase-space suppressed. The details of DM cross-section and relic can be found in Appendix \ref{appen2}. We identify two benchmark points: one for fermion (BM1) and one for scalar (BM2) DM consistent with the observed relic and experimental constraints. The details of these benchmark points are given in table \ref{tab:BM_DM}. Due to the large VEV of $\eta$ at freeze-out temperature $T_f$, a large annihilation rate of DM leads to the required relic. However, the VEV of $\eta$ relaxes to a vanishingly small value at present temperature $T_0$, consistent with the experimental constraints from direct and indirect-detection experiments. We also find the effect of entropy dilution at the end of the VEV restoring phase transition to be negligible on freeze-out relic of DM. It should be noted that our mechanism works even if the VEV of $\eta$ relaxes to the present value continuously via a second-order phase transition. However, the second step of such a two-step phase transition can be of first order generically due to the existence of a potential barrier even at tree level along the field direction \cite{Hammerschmitt:1994fn, Patel:2012pi, Inoue:2015pza, Blinov:2015sna, Ramsey-Musolf:2017tgh, Niemi:2020hto, Ghorbani:2020xqv, Benincasa:2022elt, Cao:2022ocg}.

%
%
\begin{table}[h]
    \centering
    \begin{tabular}{ccccccc}
     \hline  & $m_{\rm DM}/\mathrm{GeV}$ & $\Lambda/\mathrm{GeV}$ & $\frac{\langle \eta \rangle}{\Lambda} (T_f)$ & $\frac{\langle \eta \rangle}{\Lambda} (T_0)$ \\ \hline
      BM1 & 58 & 2045 & 0.063  & $1.1\times10^{-4}$
      \\ \hline
      BM2 & 300 & 1620 & 0.41 & $1.4\times10^{-4}$
      \\ \hline
    \end{tabular}
    \caption{Benchmark parameters used in the numerical analysis.}
    \label{tab:BM_DM}
\end{table}
\begin{figure*}
    \centering
    \includegraphics[scale=0.85]{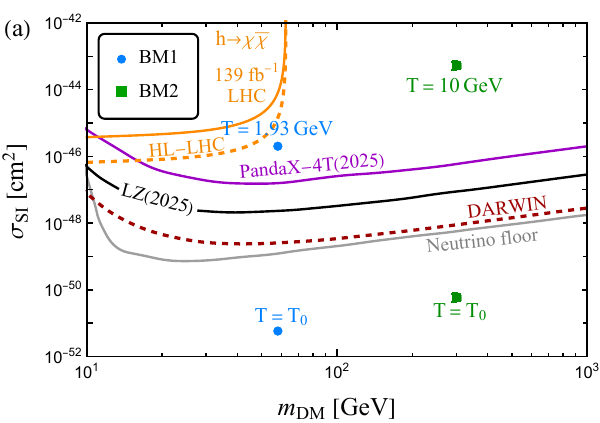} \ \ \ 
    \includegraphics[scale=0.85]{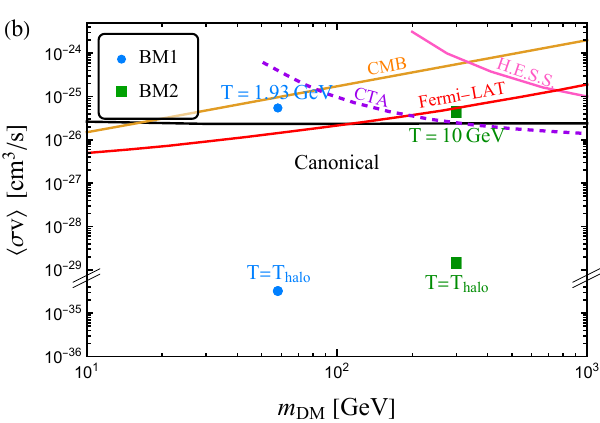} 
    \caption{Spin-independent DM-nucleon cross-section (panel (a)) and DM annihilation cross-section (panel (b)) as functions of DM mass. The points denoted as BM1, BM2 indicate the corresponding rates for fermion and scalar DM respectively considering effective DM-SM couplings at two different temperatures. Solid and dashed contours correspond to different exclusion lines and future sensitivities respectively, see text for details.}
    \label{fig:sigmaSI}
\end{figure*}

Fermion DM can lead to observable DM-nucleon scattering mediated by the SM Higgs. The corresponding cross-section is given by
\begin{equation}
   \sigma^{\chi}_{\rm SI} \approx \frac{(v'_\eta)^2}{\Lambda^4} \frac{4m_n^2 \mu^2_r }{\pi  m^4_h} f^2_n, 
\end{equation}
where $m_h$ is the SM Higgs mass, $f_n \approx 0.3$ \cite{Giedt:2009mr,Djouadi:2011aa} is the Higgs-nucleon coupling and $\mu_r = m_\chi  m_n/(m_\chi+m_n)$ is the DM-nucleon reduced mass. The DM-nucleon scattering cross-section for scalar DM is 
\begin{equation}
    \sigma^{\phi}_{\rm SI} \approx \big(\frac{v_\eta' v_h}{\sqrt{2}\Lambda^3}\big)^2 \frac{\mu^2_r}{\pi m^2_\phi} f^2_{\phi, n},
\end{equation}
where $\mu_r = m_\phi  m_n/(m_\phi+m_n)$ is the $\phi$-nucleon reduced mass and $f_{\phi, n}$ is the $\phi$-nucleon coupling \cite{Giedt:2009mr, Cao:2009uw}. DM can also be probed at indirect-detection experiments looking for DM annihilation into SM particles. Excess of gamma-rays, either monochromatic or diffuse, can be constrained from observations at such indirect-detection experiments. While DM annihilation to monochromatic photons is loop-suppressed, tree-level DM annihilation into different charged particles can contribute to diffuse gamma-rays which can be constrained by experimental data.

Fig. \ref{fig:sigmaSI} shows the compatibility of our proposed framework with both direct and indirect-detection constraints while satisfying the criteria for observed DM relic. The panel (a) shows the comparison of direct-detection rates for effective DM-SM couplings at $T=T_0$ with the ones around DM freeze-out epoch $T=T_f$. Similar comparison for DM annihilation rate during freeze-out $T=T_f$ and during present epoch inside a DM halo $T=T_{\rm halo}$ can be seen on the panel (b) by considering specific final states like third generation quarks. While DM annihilation at $T=T_f$ is dominated by $2 \rightarrow 2$ processes relying on $\langle \eta \rangle$, the ones at $T=T_{\rm halo}$ are dominated by processes independent of $\langle \eta \rangle$ such as $\chi \bar{\chi} \rightarrow \eta h, \phi \phi \rightarrow \eta f \bar{f}$. Clearly, the temperature-dependent couplings of DM keep the direct and indirect-detection rates highly suppressed at different experiments while being consistent with the thermal relic criteria due to larger DM-SM coupling at higher temperature. The black solid line on the panel (b) denotes the canonical DM annihilation cross-section required to satisfy the relic criteria \cite{Steigman:2012nb}. The annihilation cross-section of DM during the freeze-out epoch remains very close to this canonical value, as seen from the panel (b). The solid purple and black colored contours on the panel (a) correspond to upper limits on spin-independent DM-nucleon scattering cross-section from \texttt{PandaX-4T} \cite{PandaX:2024qfu} and \texttt{LZ} ~\cite{LZ:2024zvo} experiments respectively while the dashed maroon colored contour indicates the future sensitivity from \texttt{DARWIN} \cite{DARWIN:2016hyl}. The gray colored solid line corresponds to neutrino floor, below which coherent neutrino-nucleus scattering will dominate, making it difficult to distinguish it from DM induced recoil. The solid and dashed orange colored contours correspond, respectively, to the current LHC \cite{ATLAS:2023tkt} limit and future HL-LHC sensitivity \cite{Cepeda:2019klc} to the Higgs invisible decay branching ratio applicable for Higgs portal fermion DM scenario discussed in this work, considering $ h \rightarrow \chi \bar{\chi}$. Similarly, on the panel (b), the solid orange colored contour corresponds to cosmic microwave background (CMB) bound on DM annihilation rate to bottom quarks \cite{Planck:2018vyg}. The solid pink and red colored contours correspond to the gamma-ray bounds from \texttt{H.E.S.S.} \cite{HESS:2022ygk} and \texttt{Fermi-LAT} \cite{Fermi-LAT:2015att} respectively, assuming third-generation quark final states while the dashed purple line indicates \texttt{CTA}~\cite{CTA:2020qlo} sensitivity. Thus, Fig. \ref{fig:sigmaSI} clearly shows the compatibility of our framework with experimental constraints while also indicating the difficulty in satisfying these constraints if WIMP couplings at low temperature remain the same as the ones at high temperature. 

%
\begin{figure*}
    \centering
    \includegraphics[scale=0.7]{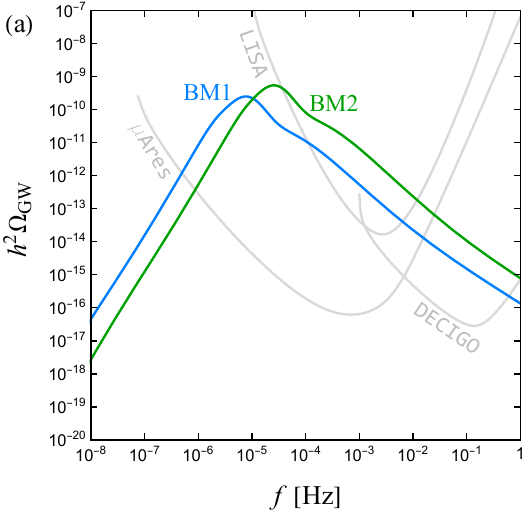} \ \ \ \ \ 
    \includegraphics[scale=0.72]{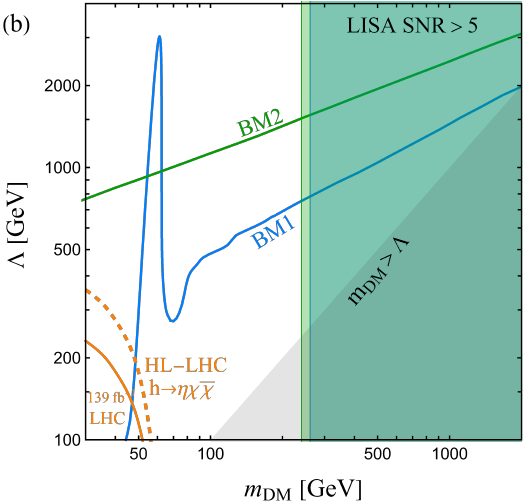} 
    \caption{Panel (a): Gravitational wave spectra for the benchmark points of fermion and scalar DM scenarios. Panel (b): $\Lambda-m_{\rm DM}$ parameter space of the model. The solid blue and green contours correspond to the points consistent with DM relic for benchmark points BM1, BM2 respectively. The LISA $\rm{SNR}>5$ region is shown in light blue and green for BM1 and BM2 respectively. The whole mass range lies in the $\mu$ARES sensitivity, which is not shown explicitly. The EFT description is invalid in the gray shaded region. The solid and dashed orange contours correspond, respectively, to current LHC limit and future HL-LHC sensitivity related to the Higgs invisible decay.}
    \label{fig:GW}
\end{figure*}

\begin{table}[h]
    \centering
    \begin{tabular}{ccccc}
     \hline  & $\alpha_*$ & $\beta/\mathcal{H}$ & $T_n/ \rm{GeV}$ & $v_w$ \\ \hline
      BM1 & $\ 3.8\ $ & $97$ & $0.9$ & $1$  \\ \hline
      BM2 & $\ 1.8\ $ & $50$ & $6$ & $1$   \\ \hline
    \end{tabular}
    \caption{Phase transition related parameters for the chosen benchmark points.}
    \label{tab:GW}
\end{table} 

{\it Experimental signatures}: While DM detection rates at ongoing or near-future experiments remain suppressed in our framework, there exist complementary detection avenues, particularly in terms of stochastic gravitational waves. Strong first-order phase transition has been studied extensively in the context of GW produced from bubble collisions \cite{Turner:1990rc,Kosowsky:1991ua,Kosowsky:1992rz,Kosowsky:1992vn,Turner:1992tz}, the sound wave of the plasma~\cite{Hindmarsh:2013xza,Giblin:2014qia,Hindmarsh:2015qta,Hindmarsh:2017gnf} and the turbulence of the plasma~\cite{Kamionkowski:1993fg,Kosowsky:2001xp,Caprini:2006jb,Gogoberidze:2007an,Caprini:2009yp,Niksa:2018ofa}. Recently, multi-step phase transitions \cite{Hammerschmitt:1994fn, Patel:2012pi, Inoue:2015pza, Blinov:2015sna, Ramsey-Musolf:2017tgh, Niemi:2020hto, Ghorbani:2020xqv, Benincasa:2022elt, Cao:2022ocg} have gained significant interest due to the possibility of realizing a strong FOPT naturally. This possibility arises due to the existence of a strong potential barrier in the field space, particularly during the second step of the transition. The key parameters, namely, the latent heat released $(\alpha_*)$, inverse duration of the transition $(\beta/\mathcal{H})$, nucleation temperature $(T_n)$ and bubble wall speed $(v_w)$ for the chosen benchmark points are given in table \ref{tab:GW}.

Fig. \ref{fig:GW} (a) shows the GW spectra in our framework considering the two benchmark points BM1, BM2 for fermion and scalar DM respectively. The gray colored contours correspond to the power law integrated sensitivities \cite{Schmitz:2020syl} of future GW detectors DECIGO~\cite{Kawamura:2006up},  LISA~\cite{2017arXiv170200786A} and $\mu$ARES \cite{Sesana:2019vho}. Due to the strong second step of the phase transition responsible for restoring the VEV of $\eta$, we get an enhanced GW amplitude as seen in Fig. \ref{fig:GW} (a). This is the primary experimental signature of our framework. Since the VEV of $\eta$ is crucially related to DM phenomenology as well as the phase transition or the nucleation temperature $T_n$, it is possible to find the range of DM mass which can be probed at GW experiments. As DM freeze-out is expected to occur before the nucleation temperature of the phase transition that is, $T_n<T_f\sim m_{\rm DM}/30$, one can use the unitarity upper limit on thermal DM mass $m_{\rm DM} \lesssim \mathcal{O}(100\, \rm TeV)$ \cite{Griest:1989wd} and lower bound on nucleation temperature $T_n \geq \mathcal{O}(\rm MeV)$ from successful BBN \cite{Bai:2021ibt}, to find the range of DM mass $\mathcal{O}(10 \, \rm MeV) \leq m_{\rm DM} \leq \mathcal{O}(100 \, \rm TeV)$ where our proposed framework is valid. This also keeps the the nucleation temperature in the range $\mathcal{O}(1 \, \rm MeV) \leq T_n \leq \mathcal{O}(10 \, \rm TeV)$ keeping the GW peak frequencies in $ \lesssim 0.1$ Hz ballpark.

Fig. \ref{fig:GW} (b) shows a summary of the model parameter space in $\Lambda-m_{\rm DM}$ plane. The solid blue and green contours indicate the parameter space consistent with the observed DM relic for benchmark points BM1, BM2 respectively. The Higgs portal coupling of DM in BM1 leads to the resonance near $m_{\rm DM} \approx m_h/2$. The entire plane is consistent with direct and indirect-detection bounds due to the vanishingly small DM-SM coupling at low temperature. The gray shaded region is disfavored as the EFT description is invalid for $m_{\rm DM} > \Lambda$. The region towards the left of the solid orange contour is ruled out by LHC limits on the Higgs invisible decay while the dashed orange contour indicates future HL-LHC sensitivity. The blue and green shaded regions correspond to the parameter space within reach of future GW experiment LISA. We have set the signal-to-noise ratio (SNR) $> 5$ for this region assuming experimental run of one year. To generate this figure, we use a common rescaling factor $m_{\rm{DM}}\rightarrow r m_{\rm{DM}}$, $\mu_{\sigma,\eta}\rightarrow r\mu_{\sigma,\eta}$, and accordingly calculated $T_n, \alpha_*, \beta/\mathcal{H}, v_w$ while keeping the dimensionless parameters of the scalar potential same as BM1 and BM2.. Since the second step of the phase transition $(v_\eta, 0) \rightarrow (v'_\eta, v_\sigma)$ occurs after DM freeze-out, the scale of DM mass can be related to the phase transition scale and the corresponding peak frequency of the GW spectrum. While we have kept the singlet scalar couplings with the SM Higgs to be negligible in our analysis, they can mix with the latter due to small but non-zero scalar portal couplings. This can lead to additional signatures at colliders or other low energy experiments. For example, the mass of the scalar $\eta$ is $0.12~\rm{GeV}$ and $0.72~\rm{GeV}$ for BM1 and BM2 respectively. If it mixes with the SM Higgs, the mixing angle $\theta$ would be bounded most strongly from meson decay constraints $\theta \lesssim 10^{-4}$  \cite{Cohen:2008nb, Winkler:2018qyg,Clarke:2013aya}. Such light scalars with tiny mixing with the SM Higgs, on the other hand, can have serious cosmological consequences if long-lived. We have checked that for $\theta>1.6\times10^{-5}$ (BM1) and $\theta>2\times10^{-8}$ (BM2), the lifetime of $\eta$ can be shorter than the BBN time scale and hence remain consistent with the observed light nuclei abundance \cite{Fradette:2017sdd}.


{\it Conclusion}: We have proposed a novel framework to revive minimal WIMP dark matter scenarios facing tight constraints from null results at direct-detection experiments. The core part of the proposal is the introduction of temperature-dependent coupling of DM-SM interactions. Such a dynamical coupling is achieved with an additional scalar field $\eta$ changing its vacuum expectation value while undergoing phase transitions. Dark matter mass is chosen in such a way that its freeze-out occurs during an epoch when $\eta$ VEV is relatively large leading to a typical WIMP type coupling of DM with the SM bath. After DM freeze-out, the VEV of $\eta$ shifts to zero or a vanishingly small value leading to suppressed WIMP interaction rates in the present epoch, consistent with null results at direct and indirect-detection experiments. The VEV restoration of the scalar field $\eta$ leads to a strong first-order phase transition with observable GW in near-future GW detectors like LISA. In order to show realistic examples, we adopt a model-independent approach and construct effective operators for both fermion and scalar DM interacting with the standard model fields. We outline the viability of our framework by choosing specific benchmark values of DM mass, cutoff scale $\Lambda$ as well as scalar potential parameters responsible for the VEV restoring first-order phase transition. While dark matter detection rates are negligible, consistent with null results so far, our framework can be probed at future GW experiments. The new scalar sector related to $\eta, \sigma$ can also be probed via their coupling to the SM Higgs.

{\it Acknowledgments}: We thank Tao Han, Brian Batell and Indrajit Saha for insightful discussions and comments on the manuscript. The work of D.B. is supported in part by the Science and Engineering Research Board (SERB), Government of India grant MTR/2022/000575 and the Fulbright-Nehru Academic and Professional Excellence Award 2024-25. For facilitating portions of this research, D.B. wishes to acknowledge the Center for Theoretical Underground Physics and Related Areas (CETUP*), The Institute for Underground Science at Sanford Underground Research Facility (SURF), and the South Dakota Science and Technology Authority for hospitality and financial support, as well as for providing a stimulating environment.

\appendix

\section{An explicit UV completion}
\label{appen3}
The Higgs portal fermion DM operator $\frac{1}{\Lambda^2} \eta \bar{\chi}\chi H^\dagger H$ can have a possible UV completion with the inclusion of a heavy singlet scalar $\xi$ \cite{Lopez-Honorez:2012tov} which interacts as
\begin{equation}
    -\mathcal{L} \supset y_\chi \bar{\chi} \chi \xi + \lambda_{12} \xi \eta H^\dagger H.
\end{equation}
Keeping other interactions involving $\chi, \eta, \xi$ to be sub-dominant compared to the above, we can identify $\frac{1}{\Lambda^2} \approx \frac{y_\chi \lambda_{12}}{m^2_\xi}$ after integrating out heavy scalar $\xi$. For validity of the EFT description $m_\xi > 2 m_\chi$ \cite{Lopez-Honorez:2012tov, Beniwal:2015sdl} and using the perturbative upper bound on Yukawa and quartic couplings, we get $\frac{1}{\Lambda} <\frac{(4\pi)^{3/4}}{2m_\chi}$. Similarly, the scalar DM operator $\frac{1}{\Lambda^3} \eta \phi \phi H \bar{f} f$ can be realized as low energy description of a UV complete theory having two additional scalar fields: a singlet $\zeta$ and a doublet $H_2$. The relevant interactions can be written as
\begin{equation}
    -\mathcal{L} \supset \mu_{12} \phi^2 \zeta + y_2 H_2 \bar{f} f + \lambda'_{12} \zeta \eta H^\dagger_2 H.
\end{equation}
Integrating out heavy scalars $\zeta, H_2$ leads to $\frac{1}{\Lambda^3} \approx \frac{\mu_{12} \lambda'_{12} y_2}{m^2_\zeta m^2_{H_2}}$.

\section{Details of the VEV restoration}
\label{appen1}
The relevant terms of the tree level scalar potential are given by
\begin{align}
    V(\sigma, \eta) & = \frac{\mu^2_\sigma}{2} \sigma^2 + \frac{\lambda_\sigma}{4} \sigma^4 -\frac{\mu^2_\eta}{2} \eta^2 +\frac{\lambda_\eta}{4} \eta^4 \nonumber + \frac{\lambda_{\sigma\eta}}{4} \sigma^2 \eta^2 \nonumber \\
    & - \tilde{\mu} \sigma^2 \eta
\end{align}
where $\tilde{\mu}\ll\mu_{\sigma,\eta}$, and we therefore ignore the contribution from the trilinear term in the phase transition and only explicitly consider it at $T=T_0\approx 0$. We use {\tt{DRalgo}} \cite{Ekstedt:2022bff} to compute the high temperature effective potential.
\begin{figure*}
    \centering
    \includegraphics[scale=0.67]{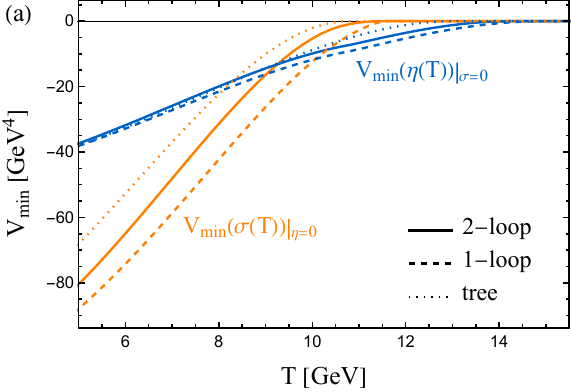}
     \includegraphics[scale=0.67]{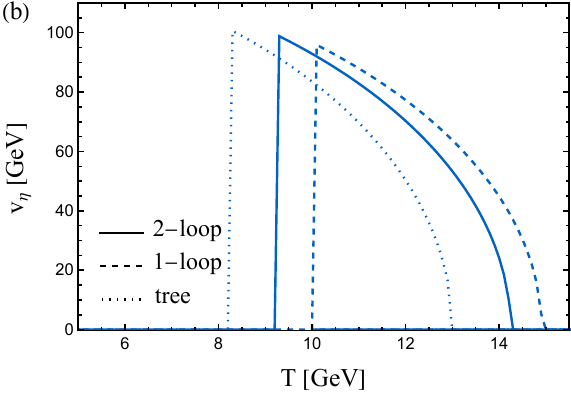} 
    \caption{Panel (a): Thermal history of the potential minima for benchmark point BM1. Panel (b): The VEV profile of $\eta$ for benchmark point BM1.}
    \label{fig:VminT}
\end{figure*}
As described in the main text, we are interested in the two-step phase transition of the two scalar fields $(\eta,\sigma)$ such that the VEVs follow $(0,0)\to(v_{\eta},0)\to(v_{\eta}',v_{\sigma})$ as the universe cools down. At $T > T_{c1}$, the VEVs of $\sigma$ and $\eta$ lie at $(0,0)$, at $T_{c1}>T > T_{c2}$, $\eta$ firstly acquires a large VEV while $\sigma$ remains zero, at $T_{c2}>T$, $\sigma$ gets a large VEV while the VEV of $\eta$ almost vanishes up to $v_{\eta}'\propto \tilde{\mu}$ from the small trilinear term, such a VEV configuration then remains till $T=T_0$ at present epoch. 
In such a phase transition, the VEV of $\eta$ at tree level is given by
\begin{equation}
    v_\eta(T)=\begin{cases}
        0,& \hspace{0.5em}T>T_{c1} \\[1mm]
        \frac{\mu_{\eta3}(T)}{\sqrt{\lambda_{\eta3}(T)}}+\mathcal{O}(\tilde{\mu}),& \hspace{0.5em}T_{c1}\geq T\geq T_{c2} \\
        \mathcal{O}(\tilde{\mu}),& \hspace{0.5em}T_{c2}>T>T_0 \\[1mm]
        \frac{v_\phi^2 \tilde{\mu}}{\frac{\lambda_{\sigma\phi}}{2}v_\phi^2-\mu_\eta^2},& \hspace{0.5em}T=T_0
    \end{cases} \label{eq:veta}
\end{equation}

Such a two-step transition requires the following relations for the couplings \cite{Ghorbani:2020xqv}
\begin{align}
    T_{c1}\approx \frac{\mu_\eta}{\sqrt{\frac{1}{4}\lambda_\eta + \frac{1}{24}\lambda_{\sigma\eta}}} & > \frac{\mu_\sigma}{\sqrt{\frac{1}{4}\lambda_\sigma + \frac{1}{24}\lambda_{\sigma\eta}}} >  T_{c2}, \\
   \Big|V_{\sigma,\mathrm{min}}|_{\eta=0}\Big| \approx \frac{\mu_\sigma^4}{4\lambda_{\sigma}} & >  \frac{\mu_\eta^4}{4\lambda_{\eta}} \approx \Big|V_{\eta,\mathrm{min}}|_{\sigma=0}\Big|,
\end{align}
where $V_{\sigma,\mathrm{min}}|_{\eta=0}$ is the minimum of the effective potential $V(\sigma,\eta)$ in the direction of $\sigma$ when $\eta=0$. We choose the benchmark parameters given in table \ref{tab:BM_PT} to achieve such a transition.
\begin{table}[h]
    \centering
    \begin{tabular}{ccccccc}
     \hline  & $\mu_\sigma/\mathrm{GeV}$ & $\mu_\eta/\mathrm{GeV}$ & $\lambda_\sigma$ & $\lambda_\eta$ & $\lambda_{\sigma\eta}$ & $\tilde{\mu}/\mathrm{GeV}$ \\ \hline
      BM1 & 3.8 & 0.11 & $0.47$ & $7.1\times10^{-7}$ & $0.001697$ & $1\times 10^{-4}$
      \\ \hline
      BM2 & 18 & $0.6$ & $0.27$ & $8.1\times10^{-7}$ & $0.001461$ & $1\times 10^{-4}$
      \\ \hline
    \end{tabular}
    \caption{Relevant parameters of the scalar potential for the benchmark points BM1 and BM2.}
    \label{tab:BM_PT}
\end{table}
In Fig.~\ref{fig:VminT} (a), we show the minimum of the effective potential along the $\sigma$ and $\eta$ axes respectively as a function of $T$ for BM1, whichever is lower is the true minimum of the potential and the intersection of which marks the critical temperature $T_{c2}$. The $\eta$ field therefore temporarily acquires a large VEV, as shown in Fig.~\ref{fig:VminT} (b), which temporarily turns on efficient DM annihilation via $2 \rightarrow 2$ processes.

\begin{figure}
    \centering
    \includegraphics[scale=0.85]{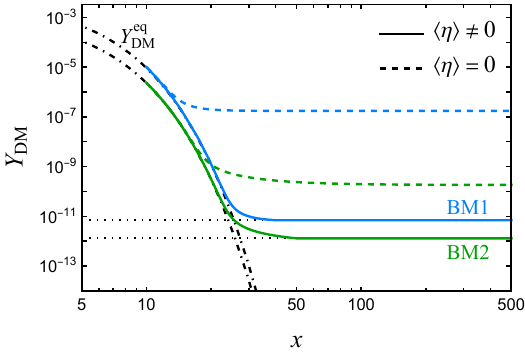} 
    \caption{Evolution of comoving DM density for fermion (BM1) and scalar (BM2) DM. The solid (dashed) lines correspond to non-zero (vanishing) VEV of the scalar field $\eta$. The horizontal dotted lines correspond to the respective relic required to fit the observed data. }
    \label{fig:Yx}
\end{figure}

\section{Details of DM relic estimates}
\label{appen2}
The relic abundance of DM can be estimated by solving the Boltzmann equation
\begin{equation}
    \frac{dY_{\rm DM}}{dx}=-\frac{s(m_{\rm DM})}{x^2  \mathcal{H}(m_{\rm DM})}  \langle\sigma_{\rm ann} v\rangle (Y^2_{\rm DM}-(Y^{\rm eq}_{\rm DM})^2),
\end{equation}
where $Y_{\rm DM}=n_{\rm DM}/s$ is the comoving number density of DM with $n_{\rm DM}, s$ being DM number density and entropy density of the Universe respectively. $x=m_{\rm DM}/T$ and $\mathcal{H}$ denotes the Hubble expansion parameter. $\langle\sigma_{\rm ann} v\rangle$ denotes the thermal averaged cross-section \cite{Gondolo:1990dk} for DM annihilation into lighter SM particles.

For fermion DM with $\mathcal{L}=-\frac{1}{\Lambda^2}\eta\bar{\chi}\chi H^\dagger H$, when $m_\chi\sim m_h/2$ and $v_\eta\neq0$ corresponding to BM1 with $T_{c1}>T>T_{c2}$, the dominant channels for DM annihilation are $\chi\bar{\chi}\rightarrow h \rightarrow f\bar{f}$ and $\chi\bar{\chi}\rightarrow h \rightarrow VV$. The corresponding cross-sections are given by
%
\begin{align}
   & \sigma(\chi\bar{\chi} \rightarrow f\bar{f})  = y^2_f \left(\frac{v_h v_\eta}{\Lambda^2}\right)^2 \frac{N_c}{16 \pi s [(s-m^2_h)^2+m^2_h \Gamma^2_h ]} \nonumber \\
   & \times  \sqrt{s-4m^2_\chi}  (s-4m^2_f)^{\frac{3}{2}}, \nonumber \\
   &  \sigma(\chi\bar{\chi} \rightarrow VV)  = \delta_V \left(\frac{v_h v_\eta}{\Lambda^2}\right)^2 \frac{1}{32\pi v_h^2}\frac{\sqrt{s-m_\chi^2}s^{3/2}}{(s-m^2_h)^2+m^2_h \Gamma^2_h }\frac{1}{\pi^2} \nonumber \\
  & \times \int\! \dd q_1^2 \dd q_2^2 \frac{m_V^2 \Gamma_V^2}{|D(q_1^2)|^2 |D(q_2^2)|^2} \sqrt{\lambda(q_1^2,q_2^2,s)} \nonumber \\
  & \times \left(\lambda(q_1^2,q_2^2,s)+ 12\frac{q_1^2q_2^2}{s^2}\right) \nonumber
\end{align}
%
where $D(q^2)=q^2-m_V^2+im_V \Gamma_V$, $\lambda(x,y,z)=(1-\frac{x}{z}-\frac{y}{z})^2-4\frac{xy}{z^2}$, and $\delta_V=1$ for $W$ and $\frac{1}{2}$ for $Z$. $N_c$ denotes the color factor for final state fermion $f$.
The DM also annihilates to $\eta h$ which is independent of $v_\eta$. The cross-section is given by
\begin{equation}
    \sigma(\chi\bar{\chi}\rightarrow\eta h) = \left(\frac{v_h}{\Lambda^2}\right)^2 \frac{1}{64\pi}\sqrt{1-\frac{4m_\chi^2}{s}}\left(1-\frac{m_h^2}{s}\right)
\end{equation}
where we ignored the mass of $\eta$.
When $m_\chi \gtrsim m_h$, DM can also annihilate through $\chi\bar{\chi}\rightarrow hh$, with three contributions, the $s$-channel $\chi\bar{\chi} \rightarrow h \rightarrow hh$ diagram, $t$ and $u$-channel diagram from $\bar{\chi}\chi H$ vertex, and contact diagram from $\chi\bar{\chi} H^\dagger H$ vertex. The cross-section is given by
\begin{align}
 &   \sigma(\chi\bar{\chi}\rightarrow hh) = \frac{\sqrt{(s-4m_h^2)(s-4m_\chi^2)}}{64\pi s} \nonumber \\
& \times \left ( A^2+ \frac{16 B^2}{s-4m_h^2+\frac{m_h^4}{m_\chi^2}}  \right ) \nonumber \\
& - \frac{m_\chi AB}{8\pi s}\ln\!\left( \frac{s-2m_h^2-\sqrt{(s-4m_h^2)(s-4m_\chi^2)}}{s-2m_h^2+\sqrt{(s-4m_h^2)(s-4m_\chi^2)}} \right) \nonumber 
\end{align}
where $A=\frac{v_\eta}{\Lambda^2}\left(1+\frac{3 m_h^2 }{s-m_h^2}\right)$, $B=\left(\frac{v_\eta v_h}{\Lambda^2}\right)^2$. 

The same interaction of fermion DM also allows Higgs invisible decay through $h\rightarrow\eta \bar{\chi} \chi$ which is independent of $v_\eta$. The width is given by
\begin{equation}
    \Gamma(h\rightarrow\eta \bar{\chi}\chi) = \frac{m_h^3}{(4\pi)^3} \left( \frac{v_h}{\Lambda^2} \right)^2 f\Big(\sqrt{1-\frac{4m_\chi^2}{m_h^2}}\Big)
\end{equation}
where
\begin{align}
    f(x)=\,&\frac{1}{12}x - \frac{1}{3} x(1-x^2)- \frac{1}{16}x(1-x^2)^2  \notag \\ 
    &  + \frac{1}{32}(1-x^2)^2(x^2+5) \ln\!\big( \frac{1+x}{1-x} \big) \label{eq:3body}
\end{align}
and we ignore the mass of $\eta$.

For the scalar DM with $\mathcal{L}=-\frac{1}{\Lambda^3} \eta \phi \phi H \bar{f} f$, when $v_\eta \neq 0$, the dominant annihilation cross-section is
\begin{equation}
    \sigma(\phi\phi\rightarrow \bar{f}f) = \frac{N_c}{4 \pi s} \left(\frac{v_h v_\eta}{\Lambda^3}\right)^2 \frac{(s-4m_f^2)^{\frac{3}{2}}}{\sqrt{s-4m_\phi^2}}.
\end{equation}
DM also annihilates to $\eta \bar{f}f$, which is independent of $v_\eta$. The cross-section is given by
\begin{equation}
    \sigma(\phi\phi\rightarrow \eta f\bar{f}) = \frac{N_c}{(4\pi)^3}\frac{2s^{\frac{3}{2}}}{\sqrt{s-4m_\phi^2}} \left(\frac{v_h}{\Lambda^3}\right)^2 f\Big(\sqrt{1-\frac{4m_f^2}{s}}\Big)
\end{equation}
where $f(x)$ is given by Eq.~(\ref{eq:3body}) and we ignore the mass of $\eta$.

Fig. \ref{fig:Yx} shows the evolution of comoving DM density for fermion (BM1) and scalar (BM2) DM with relevant parameters fixed as given in table \ref{tab:BM_DM}. When the VEV of $\eta$ is large, the DM-SM effective coupling parametrised by $\langle \eta \rangle/\Lambda$ is sizeable, leading to the required relic. When $\langle \eta \rangle $ is zero or vanishingly small, the relic is dominantly dictated by the subleading or phase-space suppressed processes, which lead to thermally overproduced relic, as shown by the dashed lines in Fig. \ref{fig:Yx}. The evolution follows the standard WIMP thermal history.


\providecommand{\href}[2]{#2}\begingroup\raggedright\endgroup

\end{document}